\documentclass[traditabstract]{aa}

\usepackage{natbib}
\bibpunct{(}{)}{;}{a}{}{,} % to follow the A&A style

\usepackage{graphics}   % standard graphics specifications
\usepackage{graphicx}   % alternative graphics specificati.jpgons
\usepackage{epstopdf}
\usepackage{longtable}   % helps with long table options
\usepackage{url}      % for on-line citations
\usepackage{bm}        % special 'bold-math' package
\usepackage{soul}

\usepackage[varg]{txfonts}
\usepackage{pdflscape}
\usepackage{supertabular}
\usepackage{hyperref}
\usepackage{xcolor}
\usepackage[normalem]{ulem}
\usepackage{natbib}
\usepackage{amsmath}
\definecolor{green}{rgb}{0.3,0.7,0.}
\definecolor{purple}{rgb}{0.77, 0.29, 0.55}
\newcommand\Mpy{\ensuremath{\ \msun\,{\rm yr}^{-1}}}
\newcommand\dm{\ensuremath{\rm{\dot{M}}}}
\newcommand\gva{{\sc Genec}}
\newcommand\stl{{\sc Stellar}}
\newcommand{\msolar} {$\rm{M_{\odot}}~$}
\newcommand{\msolarc} {$\rm{M_{\odot}}$}
\newcommand{\msolaryr} {$\rm{M_{\odot}~yr^{-1}}~$}
\newcommand{\msolaryrc} {$\rm{M_{\odot}~yr^{-1}}$}
\newcommand{\molH} {$\rm{H_2}$~}

\hypersetup{
    bookmarks=true,         % show bookmarks bar?
    unicode=false,          % non-Latin characters in Acrobat?s bookmarks
    colorlinks=true,       % false: boxed links; true: colored links
    linkcolor=blue,          % color of internal links (change box color with linkbordercolor)
    citecolor=blue,        % color of links to bibliography
    filecolor=blue,      % color of file links
    urlcolor=blue           % color of external links
}

\usepackage{color}

\newcommand{\msun}{\ensuremath{\mathrm{M}_{\odot}}\xspace}

\begin{document}

\title{Critical accretion rates for rapidly growing massive Population III stars}
\titlerunning{Critical accretion for Pop III stars}

\author{Devesh Nandal\inst{1}, John A. Regan\inst{2}, Tyrone E. Woods\inst{3},  Eoin Farrell\inst{1}, Sylvia Ekström\inst{1}, Georges Meynet\inst{1}}
\authorrunning{Nandal et al.}

\institute{D\'epartement d'Astronomie, Universit\'e de Gen\`eve, Chemin Pegasi 51, CH-1290 Versoix, Switzerland \and
Centre for Astrophysics and Space Sciences Maynooth, Department of Theoretical Physics, Maynooth University, Maynooth, Ireland \and
National Research Council of Canada, Herzberg Astronomy \& Astrophysics,
5071 West Saanich Road, Victoria, BC V9E 2E7, Canada}

\date{}

\abstract{ 
Efforts to understand the origin and growth of massive black holes observed in the early Universe have spurred a strong interest in the evolution and fate of rapidly-accreting primordial (metal-free) stars. Here, we investigate the evolution of such Population III stars under variable accretion rates, focusing on the thermal response and stellar structure, the impact of the luminosity wave encountered early in the pre-main sequence phase, and the influence of accretion on their subsequent evolution. We employ the Geneva stellar evolution code and simulate ten models with varying accretion histories, covering a final mass range from 491 \msolar to 6127 \msolarc. Our findings indicate that the critical accretion rate delineating the red and blue supergiant regimes during the pre-main sequence evolution is approximately $2.5\times10^{-2} \Mpy$. Once core hydrogen burning commences, the value of this critical accretion rate drops to $7.0\times10^{-3}\Mpy$.  Moreover, we also confirm that the Kelvin-Helmholtz timescale in the outer surface layers is the more relevant timescale for determining the transition between red and blue phases.
Regarding the luminosity wave, we find that it affects only the early pre-main sequence  phase of evolution and does not directly influence the transition between red and blue phases, which primarily depends on the accretion rate. Finally, we demonstrate that variable accretion rates significantly impact the lifetimes, surface enrichment, final mass and time spent in the red phase. 
Our study provides a comprehensive understanding of the intricate evolutionary patterns of Population III stars subjected to variable accretion rates. 
}


\maketitle

\section{Introduction}
Supermassive stars (SMSs) and  massive Population III (PopIII) stars are theorised to be a key intermediate stage in producing black holes with masses in the range $10^3$ \msolar to $10^5$ \msolar in the early Universe. Observations of distant quasars \citep{Willott_2010, Mortlock_2011, Banados_2018, Wang_2021} powered by supermassive black holes (SMBHs) with masses in excess of $10^9$ \msolar place extremely tight constraints on the time available to grow seed black holes up to these extreme masses. The recent discovery of a SMBH with a mass of approximately $10^7$ \msolar by the CEERS survey team at $z \sim 8.7$ only exacerbates the problem \citep{Larson_2023}.

While the seeds of these early SMBHs could in theory be stellar mass black holes formed from the endpoint of PopIII stars, there are a number of significant challenges to this pathway. Firstly, in order for a typical stellar mass black hole of $\approx$~100 \msolar to grow by the required 6 - 8 orders of magnitude within a few hundred Myr, it would need to accrete at the Eddington rate for the entire time.
In addition, these relatively light black holes would need to sink to the centre of their host galaxy and then find themselves continuously at the centre of a large  gas inflow to reach such masses.

A number of numerical studies have investigated this growth pathway \citep[e.g.,][]{Milosavljevic_2009, Alvarez_2009, Tanaka_2009, Jeon_2012, Smith_2018}
and in all cases have found that the stellar mass black holes struggle to achieve significant growth because they are rarely surrounded by the high density gas that is required to trigger rapid growth. Separately, a number of authors have investigated the dynamics of stellar mass black holes inside of high-z galaxies and have found that black holes with masses less than $10^5$~\msolar produce insufficient dynamical friction and do not fall to the centre of the galaxy \citep{Pfister_2019, Beckmann_2022}, instead wandering the host galaxy via a random walk
\citep{Bellovary_2010, Tremmel_2018}.

As a result of these challenges, recent efforts have focused on investigating the possibility that the early Universe may harbour the correct environmental conditions for SMS and massive PopIII star formation. True SMS formation requires that the star reaches the GR instability \citep{Chandrasekhar_1964b} due to sustained accretion up to masses of $\approx 10^5 - 10^6$~\msolar \citep{Hosokawa_2013}.
The stars we study here do not reach such high masses, so we instead refer to them as massive PopIII stars with typical masses in excess of $\approx1000$~\msolarc. The metal-free or metal-poor nature of the first galaxies is expected to provide ideal conditions that allow massive gravitationally unstable clumps of gas to form, without metal cooling lines to induce fragmentation \citet{Omukai2008}. 
Massive PopIII star formation is expected to be primarily driven by achieving the critical accretion rate needed to drive the inflation of the photosphere to become a red supergiant.
This limits radiative feedback and allows further growth of the Massive PopIII star \citep[e.g.,][]{Hosokawa_2013, Woods2017}. 
The nature and impact of this critical accretion rate is the focus of this work.

Previous studies that have investigated this problem have not identified a precise value, but it is expected to be in excess of $10^{-2}$ \msolaryrc\ \citep[e.g.,][]{Hosokawa_2013, Lionel2018}.
Such high accretion rates may be achieved in atomic cooling haloes \citep{Eisenstein_1995b, Haiman_2001, Oh_2002, Haiman_2006} or perhaps also, for short periods, during the merger of mini-haloes just below the atomic limit \citep[e.g.][]{Regan_2022}. At larger mass scales, large mass inflows may be realised during major mergers which may trigger the formation of supermassive disks \citep{Mayer_2023, Zwick_2023, Mayer_2010}.

In order to achieve this, it may also be necessary to remove, or at least strongly suppress, the abundance of \molH to avoid excessive fragmentation which should (if not limited) lower the initial mass function.
If $\lesssim 100$~\msolar PopIII stars form first, they may quickly pollute the environment with metals, cutting off the pathway to more massive PopIII star formation.  The suppression of \molH can be driven by nearby Lyman-Werner radiation sources \citep{Omukai2001b, Shang_2010, Latif_2014c, Regan_2017} which can readily dissociate \molH allowing larger Jeans masses, and may allow more massive PopIII stars to form. On the other hand, the rapid assembly of the underlying dark matter haloes \citep{Yoshida_2003a, Fernandez_2014, Latif_2022} and baryonic streaming velocities \citep{Tseliakhovich_2010, Latif_2014b, Schauer_2015, Schauer_2017, Hirano_2017, Schauer_2021} can promote massive PopIII star formation by creating conditions conducive to rapid mass inflows, which can drive massive PopIII star formation. A combinations of these processes is also possible \citep{Wise_2019, Kulkarni_2021}.

The suppression of $\lesssim 100$~\msolar PopIII star formation environments in favour of more massive PopIII star formation implies relatively very rare environments in the early Universe are required. We may note, however, that very massive PopIII stars in the early Universe have been suggested as a means to match the high luminosities observed in distant galaxies by JWST \citep{Chon_2020, Trinca_2023, Harikane_2023, Harikane_2023a}. A number of recent numerical simulations have focussed on the formation of massive objects in relatively modest Lyman-Werner radiation fields, finding that dynamical heating from major and minor mergers produces a small population of very massive (100s -- 1000s of $M_{\odot}$) stars within a parent dark matter halo \citep[e.g.,][]{Wise_2019, Regan_2020b}. These more numerous very massive stars undergo episodic rapid accretion ($\gtrsim 10^{-2}M{_\odot}$/yr) upon encountering gas overdensities within their host halo, but are otherwise quiescent, suggesting they may only occasionally sustain an inflated photosphere \citep{Regan_2020b, Woods2021}. However, the detailed evolution of very massive and supermassive stars undergoing variable rapid accretion over long timescales remains poorly understood.

A key missing ingredient in current modelling of massive PopIII star formation in cosmological settings are the transitions between inflated ``red'' and more compact ``blue'' phases. 
The exact time in the evolution of the star as well as the exact value of the accretion rate at which this occurs has been a matter of some debate in the community.
The focus of this work is to understand the evolution of rapidly-accreting massive PopIII stars with variable accretion rates drawn from the cosmological simulations of \cite{Regan_2020b}. 
This has implications for the stellar luminosity, stellar collision cross sections, radiative feedback and observable signatures of such objects. Additionally, such massive stars are expected to directly collapse into massive black holes seeding a population of intermediate mass black holes in early galaxies - possible progenitors to the recently discovered CEERS SMBH \citep{Larson_2023}.

\section{Method}

\subsection{The Stellar Accretion Rates}
The stellar accretion rates used in this research are taken from the radiation hydrodynamic simulations of \cite{Regan_2020b}. While a full discussion of these simulations is outside the scope of this article, we briefly describe the simulations and results here but direct the reader to the original paper for additional details.

The simulations undertaken by \cite{Regan_2020b} were zoom-in simulations, using the \texttt{Enzo} code \citep{Enzo_2014, Enzo_2019}, of atomic cooling haloes originally identified in \cite{Wise_2019} and \cite{Regan_2020}\footnote{The base origin of these simulations is the \textit{Renaissance} simulation suite \citep{Chen_2014, OShea_2015, Xu_2016b}}. The haloes remained both pristine (i.e. metal-free) and without previous episodes of PopIII star formation due to a combination of a mild Lyman-Werner flux and a rapid assembly history for the halo \citep{Yoshida_2003a, Fernandez_2014, Lupi_2021}. The zoom-in simulations allowed for a more in-depth and higher resolution modelling of the gravitational instabilities within the atomic cooling halo and a deeper probe of the subsequent star formation episodes which were not possible in the original$^1$, somewhat coarser resolution, simulations.

Using the higher resolution (zoom-in) simulations
\cite{Regan_2020b}, found that one of the haloes formed 101 stars during its initial burst of star formation. The total stellar mass at the end of the simulations (approximately 2 Myr after the formation of the first star) was approximately 90,000 \msolarc. The masses of the individual stars ranged from approximately 50 \msolar up to over 6000 \msolarc. The maximum spatial resolution of the 
simulations was $\Delta x \sim 1000$ au allowing for individual star formation sites to be resolved. Furthermore, accretion onto the 
stellar surface of each star was tracked and stored for the entirety of the simulation. It is these accretion rates that are used here as input to the stellar evolution modelling.

\subsection{Using the Stellar Accretion Rates as Input}

From the cosmological simulations described above, we select 10 accretion histories out of the 101 available. The choice of accretion histories was based on (i) their variability in accretion rates during the luminosity wave event and subsequent pre-MS stages, (ii) bursts in accretion rates during the hydrogen burning phases, and (iii) for a final mass range spanning over an order of magnitude. The 10 models, with variable accretion rates, are evolved from the pre-MS up until the end of core helium burning using the Geneva Stellar Evolution code (\gva) \citep{Eggenberger2008}. The models have a homogeneous chemical composition, with X = 0.7516, Y = 0.2484 and a metallicity, Z = 0 similar to \citet{Ekstrom2012} \& \cite{Murphy2021}. Deuterium, with a mass
fraction of $X_{2} = 5 \times 10^{-5}$, is also included, as in \citet{Bernasconi1996, Behrend2001} \& \cite{Lionel2018}. All models are computed with a FITM value of 0.999 (see section \ref{Sec:FITM}) in the Appendix. 
\begin{figure*}[!h]
   \centering
    \includegraphics[width=16cm]{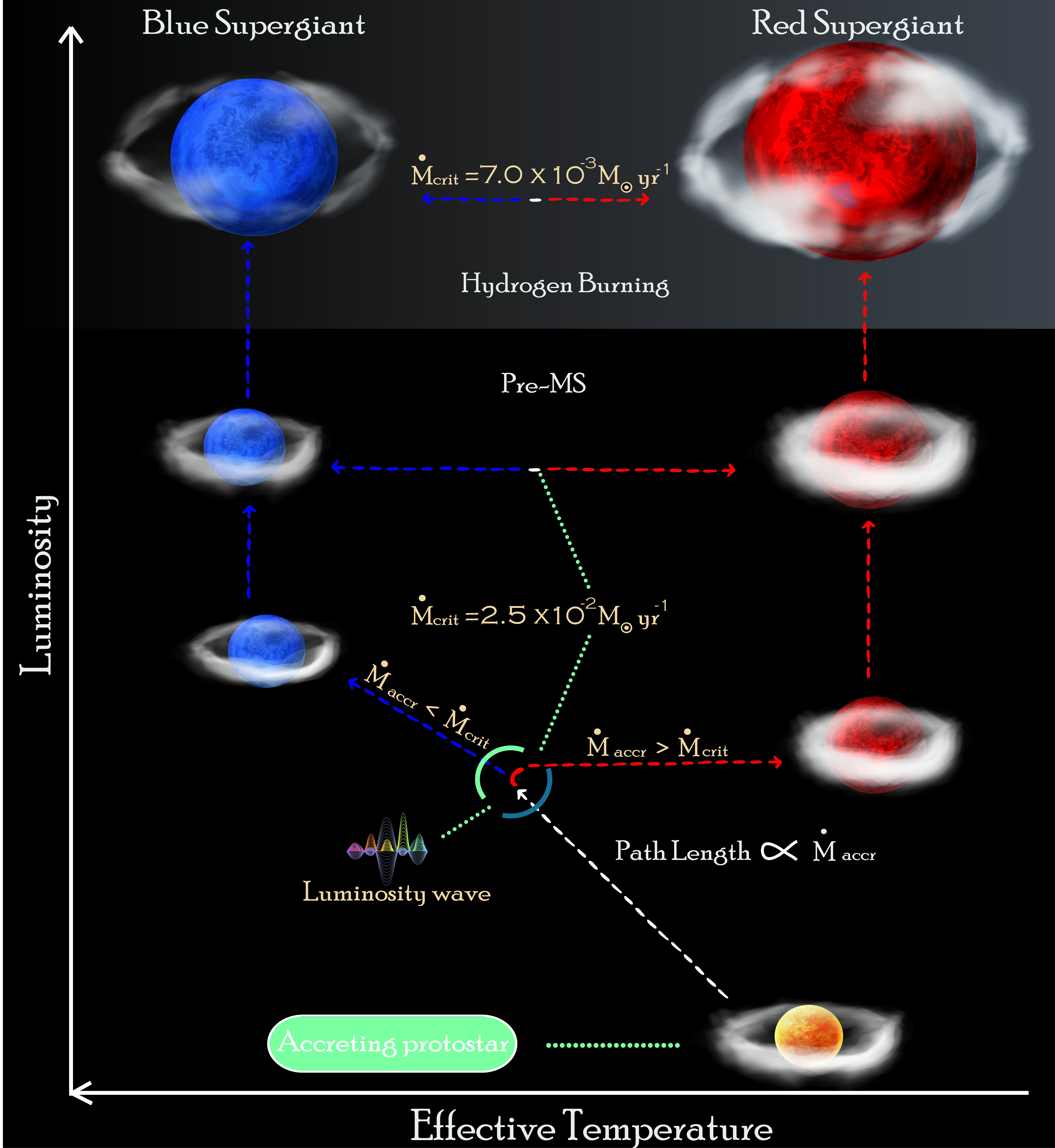}
      \caption{ An illustration depicting the impact of the critical accretion rate (\dm$_\mathrm{crit}$) on the evolution of a star. After beginning its life as a protostellar seed (in this study as a 2 \msolar seed), the star immediately migrates to hotter regions of the Hertzsprung-Russell diagram (HRD),  referred to here as the path length which is directly proportional to the accretion rate. The first transition towards a blue or red supergiant occurs when the luminosity wave breaks at the surface. In the pre-MS the evolution towards blue or red depends on a critical value of accretion rate which we determine to be 2.5$\times 10^{-2}\Mpy$. Throughout the pre-MS, if the accretion rate of a star exceeds this value, it may transition from blue to red on the surface Kelvin Helmholtz timescale; the contrary is also true. This value of the critical accretion rate is lowered to 7.0$\times 10^{-3}\Mpy$ during hydrogen burning.}
         \label{Fig:Chart_1}
   \end{figure*}

\indent Accretion commences onto low-mass hydrostatic cores with a mass of $\rm{M_{init}}$ = 2 M$_{\odot}$ for all 10 models. These initial structures correspond to n $\approx 3/2$ polytropes with flat entropy profiles, such that models begin their evolution as fully convective seeds. To model accretion, the infall of matter is assumed to occur through a geometrically thin cold disc, with the specific entropy of the accreted material being equivalent to that of the stellar surface \citep{Lionel2013, Lionel2016}. This assumption suggests that any excess entropy in the matter falling onto the star is emitted away before it reaches the stellar surface, in accordance with \cite{Palla1992} \& \cite{Hosokawa2010}. To facilitate accretion rates varying as a function of time in accordance with the hydrodynamic simulations, a new parameter that reads the accretion rates from external files was introduced into \gva. Moreover, to facilitate numerical convergence amid accretion rate fluctuations spanning 6 orders of magnitude, enhancements were made to both the spatial resolution and time discretization to accommodate this effect. Effects of rotation and mass-loss are not included in this study.

\begin{table}[h]
\caption{Physical parameters of 10 models computed from the start of pre-MS until the end of core helium burning. Column 1 shows the final mass at the end of accretion. Column 2 highlights the total duration of pre-MS evolution in kilo years whereas column 3 shows the total lifetime of models in mega years. The ratio of the time spent by models in red and blue is depicted in column 4. The mass fraction of hydrogen left in the center when models begin their final migration from blue to red in core hydrogen burning is given in column 5. The surface helium abundance at the end of computation for all models is shown in column 6 and column 7 shows the final mass of helium core.}
\label{tabModels}
\centering
\resizebox{\linewidth}{!}{
\begin{tabular}{l | c | c | c | c | c | c }
\hline
\textbf{Final Mass} & t$_{\mathrm{preMS}}$ & t$_{\mathrm{total}}$ & $t_{\mathrm{red}}/t_{\mathrm{total}}$ & X$_c$ at log  & Y$_{\mathrm{surf}}$ &  M$_{\mathrm{He}}$ \\
M$_{\odot}$ & kyrs & Myrs  & &T$_{\mathrm{eff}}=4.00$ & End He & M$_{\odot}$ \\ 
\hline \hline
$491$      &9.16&2.07& $0.08$& $0.00$& $0.40$& $114$ \\
$771$     &8.84&2.16& $0.11$& $0.00$& $0.45$& $175$\\ 
$778$     &8.60&2.03& $0.11$&$0.00$&$0.49$& $173$ \\
$932$     &9.00&1.84& $0.13$&$0.00$&$0.50$& $242$ \\
$1135$    &8.95&2.09& $0.38$&$0.27$&$0.64$& $295$ \\
$1331$   &8.70&1.86& $0.18$&$0.05$&$0.50$& $327$ \\
$1662$    &8.90&1.84& $0.59$&$0.47$&$0.76$& $662$ \\
$1923$    &9.00&1.93& $0.36$&$0.20$&$0.58$& $756$ \\
$4477$    &7.80&1.70& $0.40$&$0.23$&$0.75$& $985$ \\
$6127$    &6.86&1.51& $0.47$&$0.26$&$0.76$& $1262$ \\
\hline
\end{tabular}
}
\end{table}

\begin{figure}[!t]
	\centering
		\includegraphics[width=9cm]{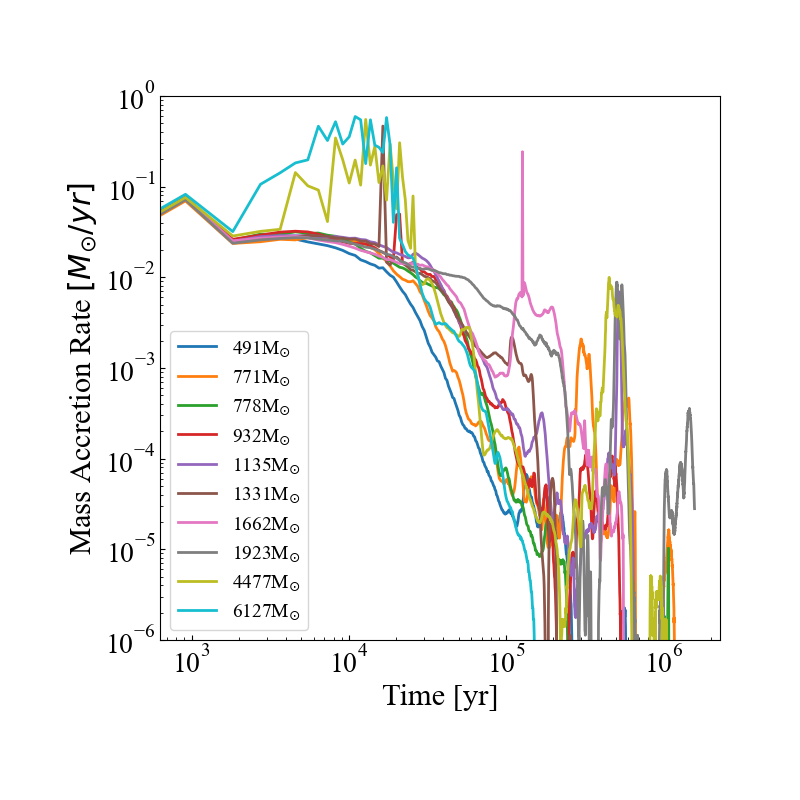}
	\caption{Accretion rate history of the 10 massive PopIII models used in this work. The data for each star is taken from \citet{Regan_2020}. The initial accretion rate is similar for each model, however, as the dynamical interaction between the stars and surrounding gas becomes dominant the stars begin to migrate outwards and away from matter rich zones and consequently the accretion rate decreases.
 }

		\label{Fig:HaloA}
\end{figure}

\begin{figure*}
	\centering
		\includegraphics[width=9cm]{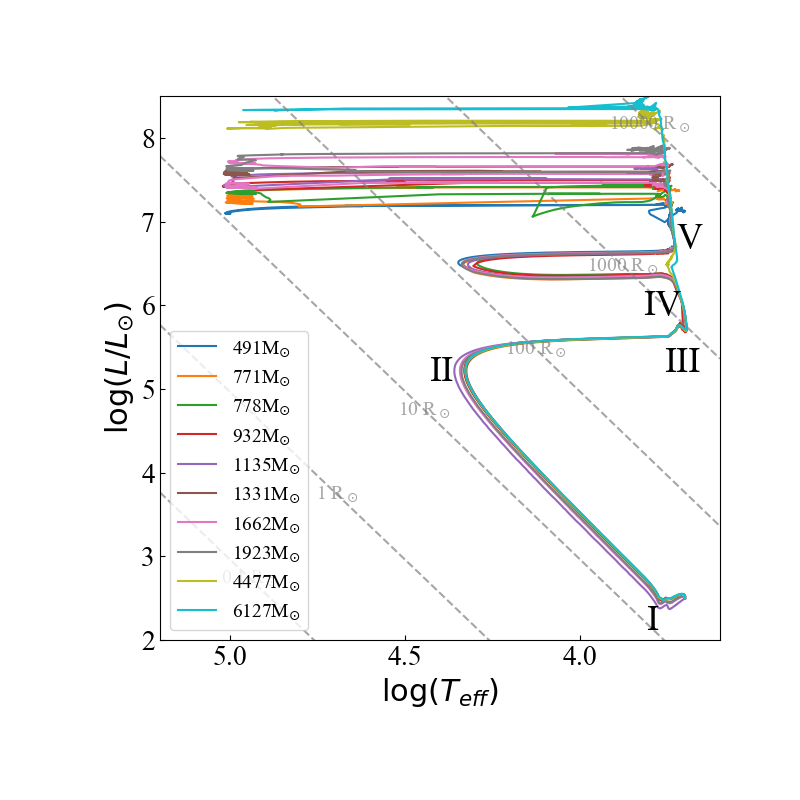}\includegraphics[width=9cm]{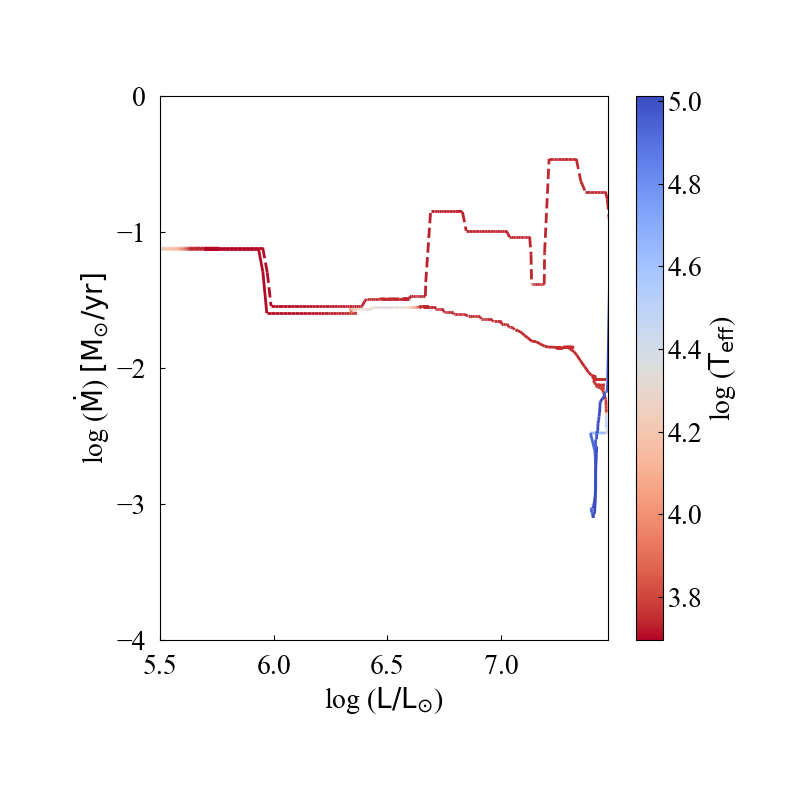}
	\caption{{\it Left panel:} Hertzberg-Russell (HR) diagram depicting the evolution for 10 models with varying accretion rates. The models are labelled by their final mass. Grey dashed lines represent isoradii. Models 491, 771, 778, 932, 1135, 1331, 1662, and 1923 have near-identical pre-MS accretion histories, unlike models 4477 and 6127. All models except 4477 complete accretion before hydrogen burning starts at log $(T_{\mathrm{eff}}) = 5.10$. Computation stops after core helium burning, with all models in the red at log $(T_{\mathrm{eff}}) \approx 3.76$.
      {\it Right panel:} The evolution of the accretion rate versus the luminosity for the 1662 (solid line) and 4477 (dashed line) models is presented and colour-coded by the effective temperature. Quantities are displayed on a log scale, focusing on a zoomed-in region of the pre-MS. Both models share an identical accretion history until reaching a luminosity of log $(L/L_\odot) = 5.95$. The 1662 model experiences a drop in the accretion rate below $2.5\times 10^{-2} M_\odot \text{ yr}^{-1}$, while the 4477 model remains above this value.}% As a result, the 1662 model evolves towards the blue, and the 4477 model stays in the red. This suggests that an accretion rate of $2.5\times 10^{-2} M_\odot \text{ yr}^{-1}$ represents the critical accretion rate during the pre-MS ($\dot{M}_{\mathrm{crit, preMS}}$).}
		\label{Fig:HRD1}
\end{figure*}

\section{Evolution of Accreting Massive PopIII Stars}\label{Sec:Evolution}
We investigate the evolution of accreting massive PopIII stars using the ten models with varying physical parameters (Table~\ref{tabModels}). The models will be referred to by their final mass; for instance the 491 \msolar star will be referred to as model 491. Critical accretion rates during the pre-main sequence (pre-MS) and core hydrogen burning phases significantly impact the stars' development. Figure~\ref{Fig:Chart_1} illustrates general trends and the dependence on the accretion rate of whether or not the star evolves to the red supergiant phase or to the blue supergiant phase. Figure \ref{Fig:HaloA} shows the accretion histories of the 10 models computed in this work. The accretion history The critical accretion rate values determined here will be explored in detail in the upcoming sections. We begin our evaluation of the critical accretion rate in the pre-MS.

\begin{figure*}[!t]
   \centering
    \includegraphics[width=9.3cm]{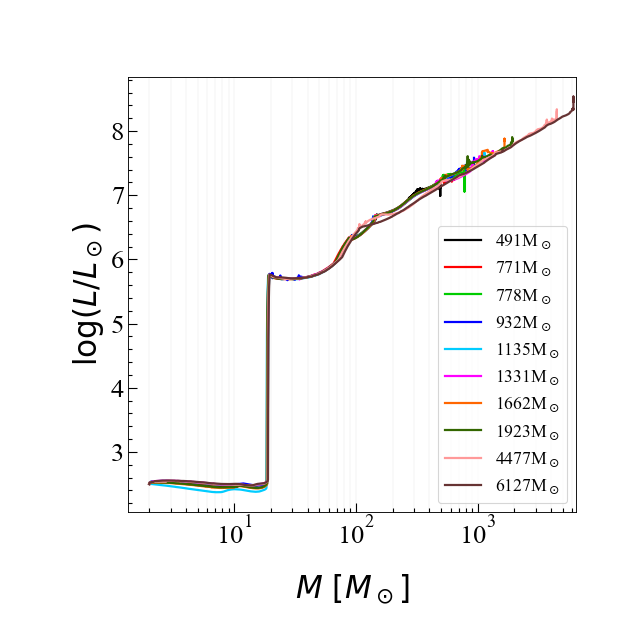}\includegraphics[width=9.3cm]{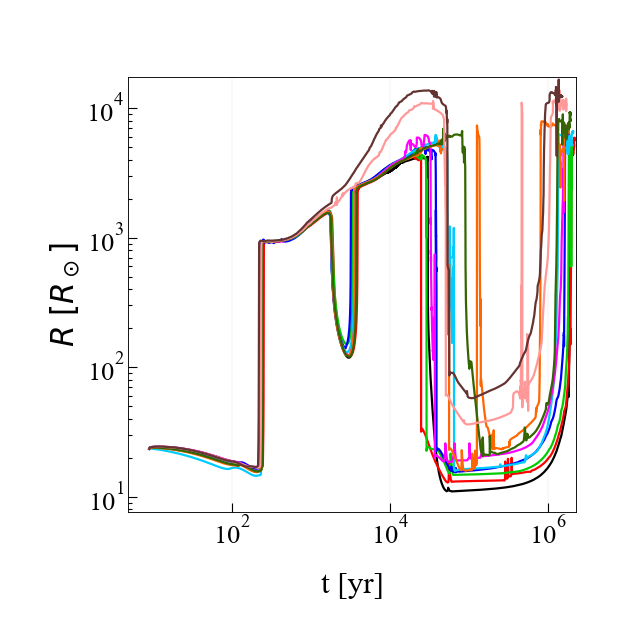}
      \caption{{\it Left panel:} Luminosity-Mass relation depicted for all models. The increase in luminosity at 20 \msolar for all models correspond to the luminosity wave breaking at the surface. Beyond this, Luminosity evolves monotonically versus mass as L $\propto$ M until the end of accretion history. 
      {\it Right panel:} Evolution of radius versus the age (in years) for all models. The colors used to represent models correspond to label shown in left panel.} %Despite models possessing unique accretion histories, the final state attained is a red supergiant star with radius ranging from 4000 R$_\odot$ to 10000 R$_\odot$. Another feature depicted is the total lifetime of star is decreases slightly as the final mass increases, a result similar to the findings of massive star models. One indeed expects that for massive star, the core H-burning phase does not much depend on the mass since, due to the effects of radiation pressure, $L \propto M$ and  the stellar lifetime $\propto M/L$ = constant.}
         \label{Fig:Lum}
\end{figure*}

\subsection{preMS evolution}\label{Sec:prems}

 The evolution onto each star commences with hydrostatic seeds having a mass of 2 M$_{\odot}$, radius of 26 R$_{\odot}$,  luminosity of $\log (L/L_{\odot})=2.50$, effective temperature of log (T$_{\rm eff}) = 3.70$, and an accretion rate of $\approx7.5\times10^{-2}\Mpy$. 
 In the left panel of Figure \ref{Fig:HRD1} we first display the paths of each star on the Hertzberg-Russell (HR) diagram and additionally indicate, with Roman numerals `I - V', events of particular note while also paying attention to the variable accretion rates. \\
 \indent I. Upon reaching $\log (L/L_{\odot})=2.47$ and log (T$_{\rm eff}) = 3.77$, all models experience the start of luminosity wave. An increase in central temperature leads to a decrease in central opacity, transitioning the convective core to a radiative core. The lowered opacity boosts luminosity production, allowing the central luminosity to migrate outwards \citep[see e.g.][]{Larson1972,Stahler1986,Hosokawa_2010,Lionel2018}. The luminosity wave breaks at the surface, with models migrating to the blue region of the HR diagram (higher effective temperatures). The path length of the knee-like feature at $\log (L/L_{\odot})=5.40$ and log (T$_{\rm eff}) = 4.35$ in the HR diagram (`I - III' in Figure~\ref{Fig:HRD1}) is directly proportional to the accretion rate during this event (for details, see also Figure \ref{Fig:Tests2} in the Appendix. The migration of luminosity wave is explored further in section \ref{Sec:Lumwave} of the Appendix). \\
\indent II. The start of this event (`II') marks the end of luminosity wave after it has travelled from center to surface over a period of $\approx 190 yr$. Following the brief period of the luminosity wave, almost all models follow near identical paths through the pre-MS. This is due to the accretion rate of all models exceeding $3.1\times10^{-2}\Mpy$. Previous works by \citet{Woods2017,Lionel2018} indicate that an accretion rate greater than $1.0\times10^{-3}\Mpy$ after the appearance of the luminosity wave results in a transition to the red. To better explore this narrow critical accretion rate regime, we performed numerical tests on models 1662 and 4477. The comparison between these two models is illustrated in the right hand panel of Figure \ref{Fig:HRD1}. By exploring the contrast between these two models in particular, we found that an accretion rate greater than $2.5\times10^{-2}\Mpy$ is needed to transition the models to the red and this value will be referred to as the critical accretion rate during the pre-MS evolution ($\dot{M}_{crit, preMS}$). The accretion timescale during this event (`II') remains nearly constant and this is due to the accretion histories we obtain from the hydrodynamic simulations. However, the surface Kelvin-Helmholtz timescale (computed at a given Lagrangian coordinate) increases, preventing the models from adjusting their structures as new matter is deposited on the surface. This results in an increase in radius and all models transition to a radiative core with a large convective envelope, becoming red supergiant protostars \citep{Hosokawa_2012}. This transition to red is extremely short and occurs over a span of 4 years, as marked by the `III' in the left panel of Figure \ref{Fig:HRD1}. After experiencing the luminosity wave and migrating to red, all of the models follow a near monotonical relationship between luminosity and mass (see left panel of Figure \ref{Fig:Lum}). This relationship of L$\propto$M is seen in all accreting models of \citet{Hosokawa_2013,Woods2017,Lionel2018} and is similar to the mass-luminosity relation for massive stars on the ZAMS (Zera Age Main Sequence). \citep{Ekstrom2012,Murphy2021}.

\indent IV. Following the models still in the left hand panel of Figure \ref{Fig:HRD1} we see that all models begin to diverge in their evolutionary path at $\log (L/L_{\odot})=5.70$ and $\log (T_{\rm eff}) = 3.70$. All models except 4477 and 6127 experience a drop in accretion rate below $\dot{M}_{crit, preMS}$ and begin their contraction towards the blue. This drop, for most models, in the accretion rate only occurs for around 800 years before again increasing in excess of $\dot{M}_{crit, preMS}$, which forces the models to migrate back to red. Models 4477 and 6127 do not follow this particular path. Instead since their accretion rates remain in excess of $\dot{M}_{crit, preMS}$ they follow the Hayashi line and remain in the red (see also Figure~\ref{Fig:Chart_1}).\\
\indent V. All ten models converge on the HR diagram at $\log (L/L_{\odot})=6.70$ and $\log (T_{\rm eff}) = 3.73$, evolving along the Hayashi line for approximately 20,000 yr. The radius of all models during this stage is between 4000 - 10,000 R$_\odot$ and is clearly shown in the right panel of Figure \ref{Fig:Lum} as the initial strong spike in the stellar radius at the end of the pre-MS. The large radius and the lack of any nuclear reaction classes such objects as red supergiant protostars \citep[see][]{Hosokawa_2013}. Additionally, the radius of each model during this stage is proportional to the current mass, which is in turn determined by the net average accretion rate prior to this stage. All models then begin their subsequent contraction towards the blue and experience a reduction in radius to $R< 60 \rm{R_\odot}$ (see again the right panel of Figure \ref{Fig:Lum}) until central conditions are optimal for core hydrogen burning. By this stage, each star is experiencing little or no accretion.

In conclusion, the luminosity wave plays a minor role in massive protostars' pre-MS evolution. However, the critical accretion rate, $2.5\times10^{-2} , \mathrm{M}_{\odot}\mathrm{yr}^{-1}$, is crucial in shaping their behavior and structure. Models with accretion rates above this value form red supergiant protostars along the Hayashi limit, while those below contract and migrate blueward. This critical accretion rate is the determining factor in understanding the massive protostars' HR diagram trajectory during the pre-MS.

\begin{figure*}
   \centering
    \includegraphics[width=18cm]{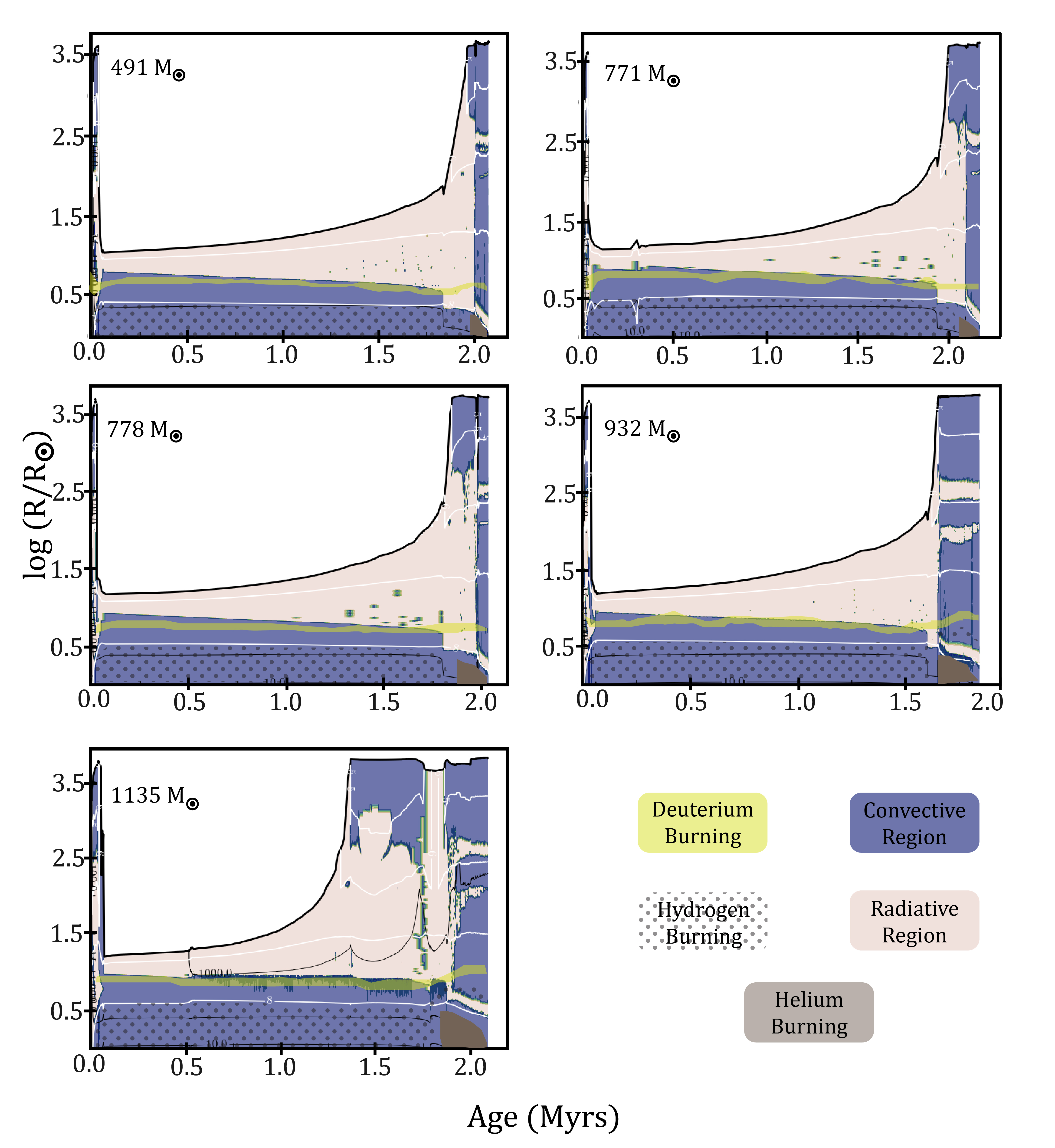}
      \caption{ Kippenhahn diagrams showing the evolution of the structure (in Eulerian coordinates) as a function of time (Myrs) for the lowest-mass models. The blue and cream regions represent the convective and radiative zones respectively. The iso-masses are depicted by black lines whereas the isotherms of log(T[K]) = 5, 6, 7 and 8 are drawn in white lines. The translucent yellow regions show deuterium burning, dotted dark grey zones are hydrogen burning zones, dark grey zone highlights helium burning. The pre-MS evolution lifetime is much shorter than the nuclear burning lifetime, therefore the leftmost section of each plot represents this stage. Here the radius of all models fluctuates over 2 orders of magnitude. The hydrogen burning phase is marked by a shrinking convective core and an inflated radiative zone. Following a short contraction phase, all models expand and transition to a core helium burning phase with a near-full convective structure.}
         \label{Fig:Kip1}
   \end{figure*}
\begin{figure*}
   \centering
    \includegraphics[width=18cm]{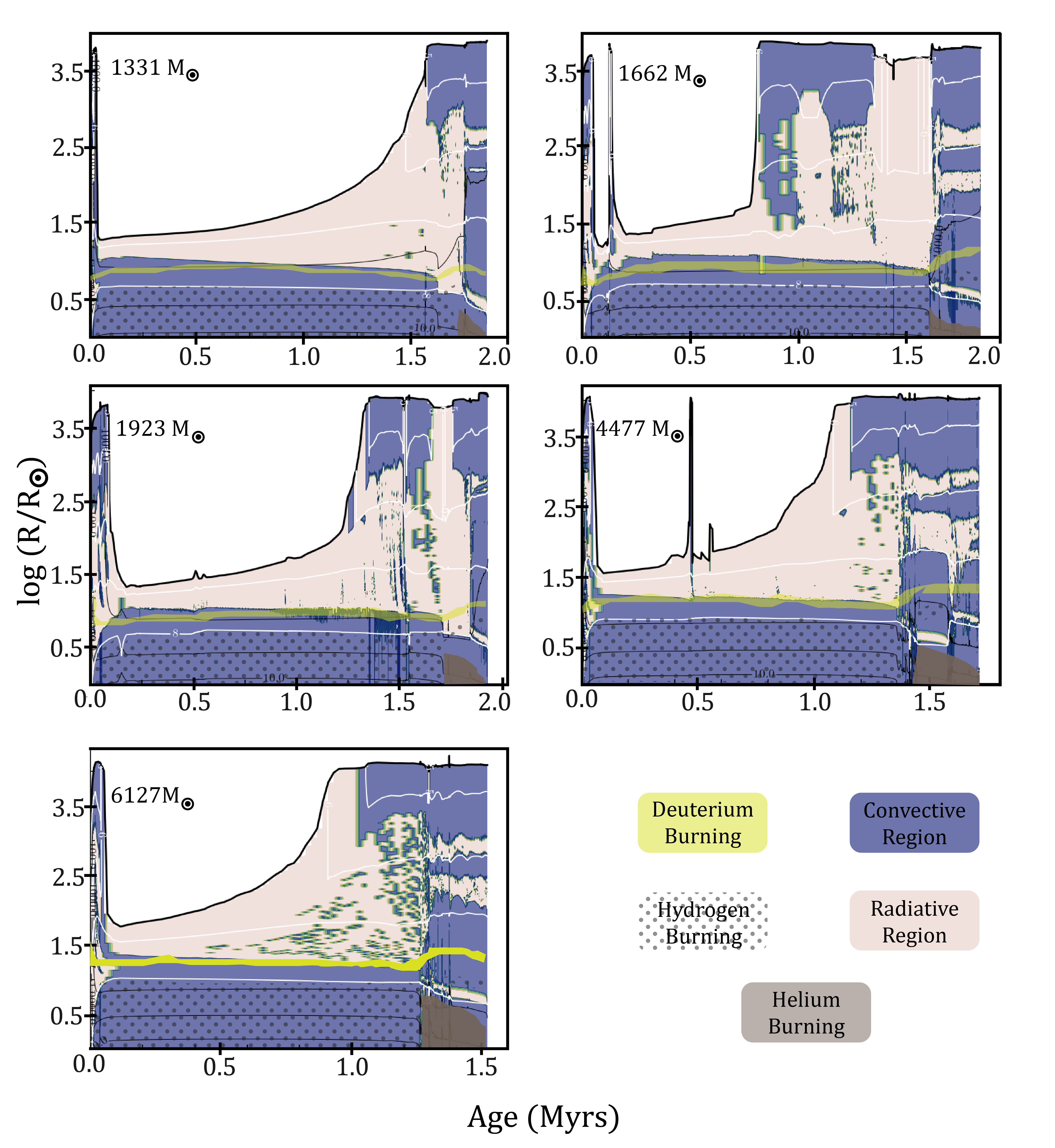}
      \caption{Kippenhahn diagrams similar to Figure~\ref{Fig:Kip1} but for models with mass 1331, 1662, 1923, 4477 and 6127 \msolarc. The radii of these models is larger than the low mass models. The models depicted here transition to the red and expand their radius before the completion of core hydrogen burning phase. Model 4477 \msolar shows an exceptional behavior as it undergoes an expansion in radius during hydrogen burning. This is due to a burst of accretion experienced by this model.}
         \label{Fig:Kip2}
   \end{figure*}

\begin{figure*}
   \centering
    \includegraphics[width=18cm]{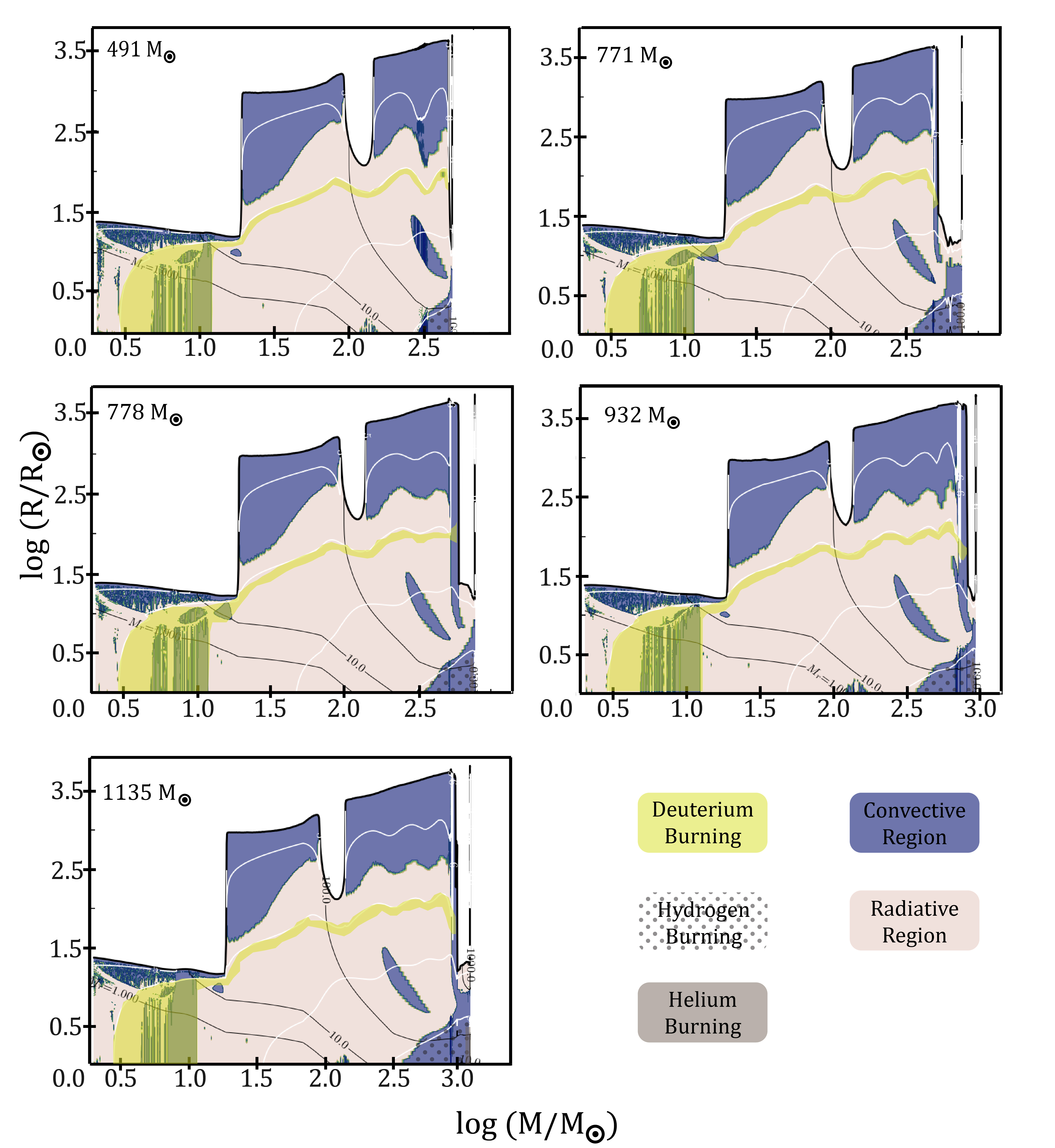}
      \caption{Kippenhahn diagrams depicting the evolution of radius versus mass (both axis represented in log scale) for low mass models. The details of diagrams are similar to Figure~\ref{Fig:Kip1}. In this version of diagrams, the change in radius during the pre-MS evolution of all models is clearly visible. All models undergo the luminosity wave at log (M/M$_{\odot}$) = 1.3 and experience a strong increase in radius. The next noteworthy evolutionary trend is at log (M/M$_{\odot}$) = 1.95 when model undergo a drop in accretion rate below \dm$_\mathrm{crit, preMS}$ which results in a decrease in radius. Eventually the accretion rate increases above the \dm$_\mathrm{crit, preMS}$}
         \label{Fig:Kip3}
   \end{figure*}

\begin{figure*}
   \centering
    \includegraphics[width=18cm]{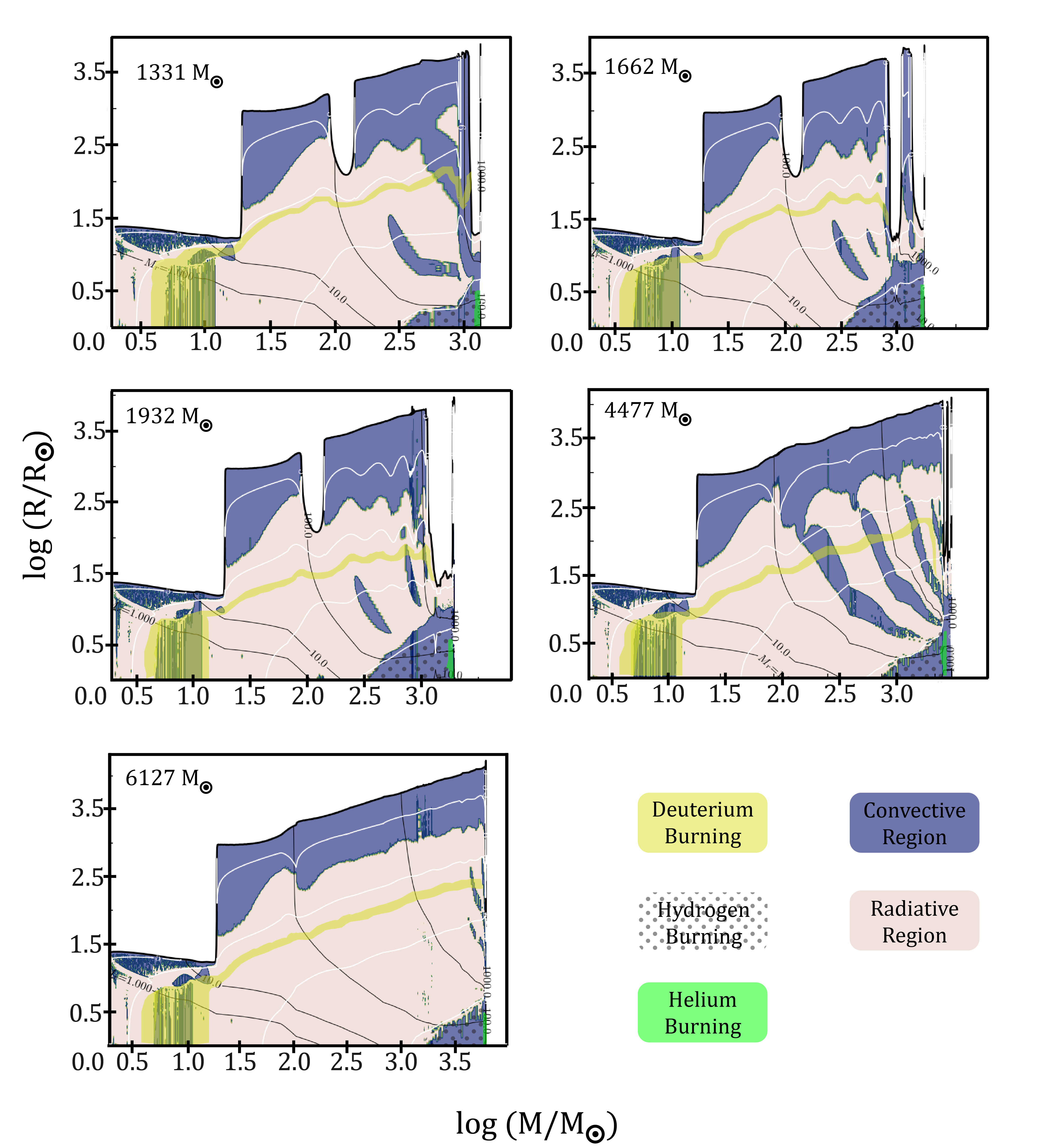}
      \caption{Kippenhahn diagrams depicting the evolution of radius versus mass (both axis represented in log scale) for high mass models with mass 1331, 1662, 1923, 4477 and 6127 \msolarc.. The details of diagrams are similar to Figure~\ref{Fig:Kip3}.}
         \label{Fig:Kip4}
   \end{figure*}

\subsection{Core Hydrogen Burning Evolution}

Core hydrogen burning commences with all models at log(T$_{\rm eff}) \approx 5.00$ and a luminosity range between $\log (L/L_{\odot})$ 7.40 and 8.30 (see left panel of Figure~\ref{Fig:HRD1}). Numerical tests indicate that for the same value of luminosity and effective temperature at ZAMS, the choice of accretion history does not impact the position of models (see section \ref{Sec:history} of Appendix).  We now examine three representative models: 491, 6127, and 4477.

Model 491 is the least massive and has no accretion during core Hydrogen burning. Hydrogen ignites in the core through pp and $3-\alpha$ reactions as the temperature exceeds $1\times 10^{8} \rm K$. The CNO cycle becomes the dominant energy source for the rest of this phase \citep{Woods2017,Lionel2018}. The top left panel of Figure~\ref{Fig:Kip1} and ~\ref{Fig:Kip3}  shows the structure, with a convective core, radiative intermediate zone, and outer convective envelope of this least massive star (491 model). The convective core mass starts at 453 M$_\odot$ and completes the main sequence in 1.82 Myrs with a final core mass of 243 M$_\odot$.

Model 6127 has the highest final mass and also stopped accreting before H ignition in the core. It starts hydrogen burning in a convective core with an initial mass of 5741 M$_\odot$ and completes the main sequence in 1.25 Myrs with a final core mass of 3593 M$_\odot$. When the effective temperature reaches log (T$_{\rm eff}) = 4.00$, the model still undergoes hydrogen burning with 0.26 X$_c$ left in the core. Core hydrogen exhaustion occurs at the Hayashi limit, followed by structural expansion and the emergence of a convective envelope with intermediate convective zones (see bottom left panel of Figure~\ref{Fig:Kip2} and ~\ref{Fig:Kip4}).

\begin{figure}[!t]
	\centering
		\includegraphics[width=9cm]{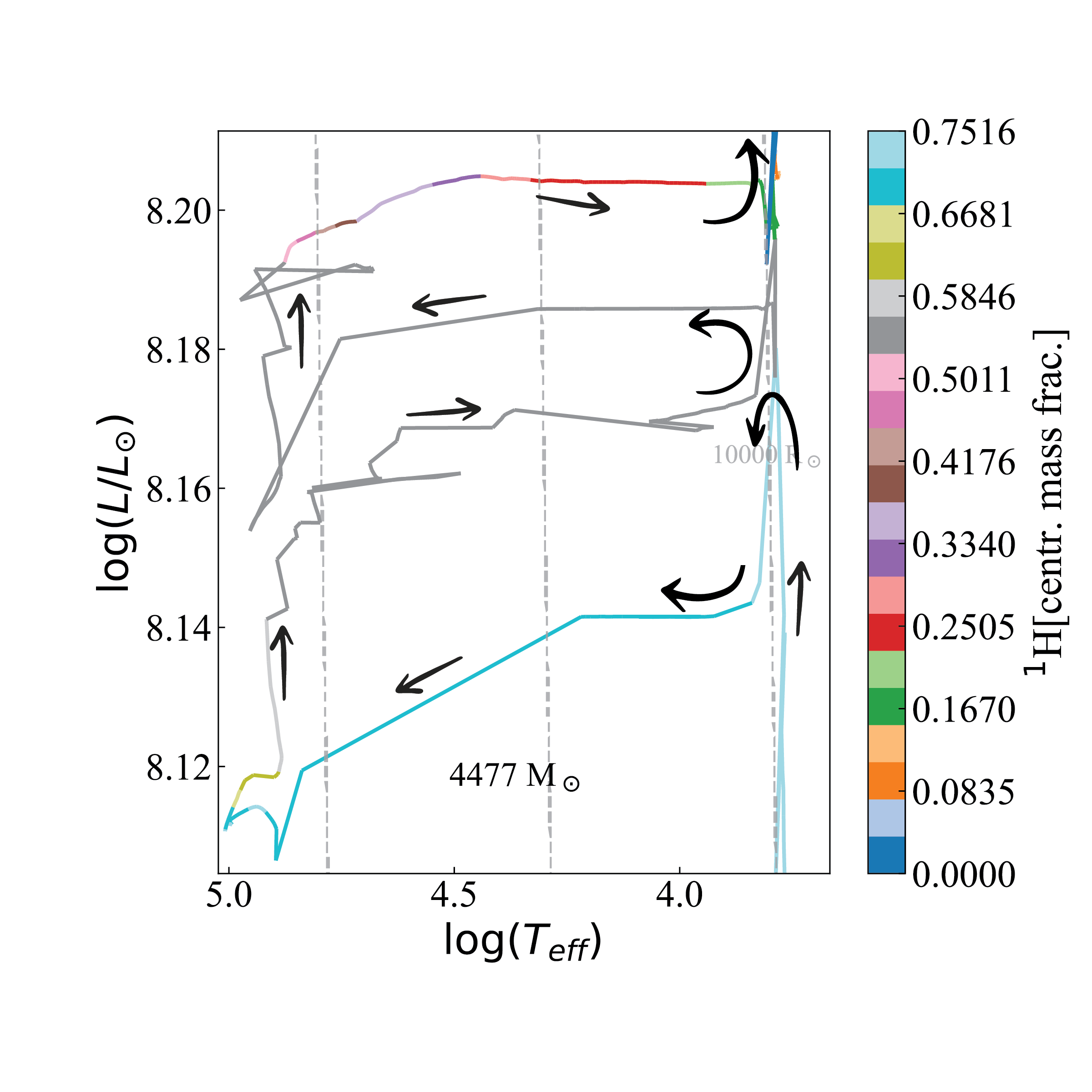}
	\caption{Magnified HR diagram of model 4477 (only the upper part is shown) depicting the migration between blue and red as result of the variable accretion rate.  The colour code follows the central mass fraction of hydrogen. The black arrows represent the evolutionary path taken by the model; the arrows emerge from bottom right which marks the pre-MS stage and end at top right which denotes the end of core hydrogen burning stage. The first migration from blue to red occurs at log (L/L$_\odot$) $\approx 8.14$ when accretion rate increases to 7.0$\times 10^{-3}\Mpy$, the new critical accretion rate during hydrogen burning. Upon reaching log (T$_{\rm eff}) \approx 3.76$, accretion rate is reduced below this critical value to 5.0$\times 10^{-3}\Mpy$. This transition from blue-red-blue occurs over a timescale 17.5 kyrs indicative of the thermal relaxation timescale during Hydrogen burning.
    }
		\label{Fig:HRD2}
\end{figure}

The 4477 model stands out from the other 10 models due to its unique accretion history extending into hydrogen burning, causing multiple blue-red transitions as shown in Figure~\ref{Fig:HRD2}. It starts hydrogen burning without accretion at $\log (T_{\rm eff})=5.00$ and $\log (L/L_{\odot})=8.11$. After just 4.0 kyr, it undergoes an accretion episode and climbs the HRD to a near-constant $\log (T_{\rm eff})=4.91$ and $\log (L/L_{\odot})=8.14$. With an accretion rate exceeding 7.0$\times 10^{-3}\Mpy$, the model is unique and migrates towards the red, revealing a critical  accretion rate that severely impacts the model's radius during hydrogen burning. This migration occurs over a Kelvin Helmholtz timescale (17.5 kyr) which is much shorter than all other models moving at nuclear timescales without accretion. Once the model finishes the redward migration, at $\log (T_{\rm eff})=3.78$ and $\log (L/L_{\odot})=8.19$, the accretion rate reduces to 5.0$\times 10^{-3}\Mpy$ after 8.5 kyr, starting its final blue-ward transition for 1.0 kyr. Arriving in the blue at $\log (L/L_{\odot})=8.15$ and $\log (T_{\rm eff})=4.94$, it exhausts all accreting matter. It then begins its final redward excursion at a nuclear timescale, lasting 0.61 Myr, ending this phase at $\log (L/L_{\odot})=8.21$ and $\log (T_{\rm eff})=3.79$. The center right panel of Figure~\ref{Fig:Kip2} and ~\ref{Fig:Kip4} shows particular trends in this phase, with a radius change at 0.45 Myr corresponding to a 25-kyr accretion rate burst. See also the right hand panel of Figure \ref{Fig:Lum}, which shows the near delta-like spike in radius of model 4477 at T $\sim$ 0.45 Myr. A large outer convective zone appears before the end of core hydrogen burning. Similar to model 6127, core hydrogen burning finishes after this model has already reached the Hayashi line.

In summary, we find a critical accretion rate of 7.0$\times 10^{-3}$ \msolaryr during the core hydrogen burning phase, as shown in model 4477. This critical accretion rate impacts the model's radius during hydrogen burning and is responsible for the unique blue-red transitions observed. It is important to note that this critical accretion rate is lower than the critical accretion rate observed during the pre-main sequence evolution.

\subsection{Helium burning}
All models follow near identical evolutionary trends during the core helium burning phase, denoted by the grey regions in Figure~\ref{Fig:Kip1} and Figure~\ref{Fig:Kip2}. By this time, accretion has completely ceased for all models and the evolution commences and ends in close proximity to the Hayashi line. To further highlight details of this stage, we will explore the least massive (model 491) and most massive (model 6127) models.

The core of model 491 undergoes a contraction until the central temperature reaches $2.80\times 10^{8} \rm K$. Helium is ignited in the convective core of mass 242 \msolarc. The external layers of the model transition from fully radiative into mostly convective zones; 85 \% of the model is now convective. Core helium burning lasts 0.23 Myrs and the final mass of the Helium core at the end of the evolution is 114 \msolarc. Model 6127 undergoes a core helium burning phase that is nearly identical to the other models, with one notable difference in its structure; it transitions into a near fully convective structure already at the end of core hydrogen burning.
A consequence of the near fully convective structure of all models during helium burning is the ability of the models to transport helium from the core to the surface, even with the absence of any rotational mixing. 
The duration of core helium burning is 0.15 Myrs and the final mass of helium core is 1262 \msolarc.
Model 491 has a surface helium abundance, Y$_{\mathrm{surf}} = 0.40$ whereas model 6127 has an extremely enriched surface, Y$_{\mathrm{surf}} = 0.76$ hinting that its value increases as a function of mass. 
Additionally, the core mass at the end of this evolutionary phase, as seen in the last column of Table \ref{tabModels} is approximately a quarter or a sixth of the total mass. 
Assuming the final fate of such objects is to form a black hole, the resulting mass would be constrained by two limits. The upper limit corresponds to the scenario where the total mass at the end of the evolution is consumed by the black hole. In this case, the black hole's mass would be equivalent to the star's final mass. Alternatively, the lower limit arises when a portion of the outer envelope situated above the helium core is lost, either due to stellar winds or instabilities occurring prior to the final collapse. Under this circumstance, the black hole's mass would be equivalent to the mass of the helium core.

\section{Discussion}\label{Sec:Discussion}

\subsection{Determining the Critical Accretion Rate}
Determining the critical accretion rate which leads to the realisation of the canonical SMS has been the goal of a number of studies over the last two decades. Seminal work by \citet{Omukai2001,Omukai2003} explored the critical accretion limit in the case of the spherical accretion of matter and found a value of $\approx 4\times10^{-3}\Mpy$. Building on this \citet{Schleicher2013} explored the relationship between the Kelvin Helmholtz timescale and the accretion timescale, and using this approach determined a mathematical expression for the critical accretion rate.
They concluded that in the case of spherical accretion, the minimum accretion rate needed to evolve a star along the Hayashi line is the somewhat higher value of $ 1\times10^{-1}\Mpy$. By exploring the more realistic case of accretion via a geometrically thin disc, \cite{Hosokawa_2012} found that the critical limit is decreased by an order of magnitude to approximately $1\times10^{-2}\Mpy$. \\
\indent \citet{Hosokawa_2013} using the \stl~ code \citep{Yorke2008} computed models and predicted the critical accretion rate for the cold disc accretion scenario. They found that with a choice of three constant accretion rates of 0.01, 0.1, and 1 $\Mpy$, the lowest accretion rate required for a star to remain in the red phase is approximately $1\times10^{-1}\Mpy$. These results were dependent on the amount of gravitational energy deposited in the center which influences the early pre-MS evolution when the Kelvin Helmholtz timescale is longer than the accretion timescale. The ratio between the accretion timescale and the Kelvin Helmholtz timescale was found to be an important metric when considering the migration towards blue or red. The choice of constant accretion rates was a limiting factor in these studies and was addressed shortly afterwards by \citet{Vorobyov2013}, when they performed 2D hydrodynamic simulations to account for a variable accretion rate. Their models underwent a burst ($\rm{\dm = 10^{-2} - 10^{-1}}\Mpy$) in accretion followed by quiescent phases ($\rm{\dm = 10^{-5} - 10^{-4}}\Mpy$) due to the migration of fragmented clumps of matter onto the star. Their models hinted at the strong impact of a varying accretion rate to the early evolution of accreting stars. Considering the impact of such accretion histories on the evolution of primordial stars, \citet{Sakurai2015} computed models with variable accretion rate and determined the critical accretion rate needed to produce a red or a blue star to be approximately $4\times10^{-2}\Mpy$.

Additionally, they found that since the mass distribution of such objects is mainly concentrated towards the center, the global Kelvin Helmholtz timescale provides a poor estimate of the overall thermal timescale. Instead, to determine whether a star would transition into a red or a blue supergiant they noted that it is important to look at the surface Kelvin Helmholtz time scale of the individual layers. 

Using \gva, \citet{Lionel2018} computed SMS models with constant accretion rates ranging from $10^{-3} - 10^{1}\Mpy$ and found that the $10^{-2}\Mpy$ model exhibited an oscillatory behaviour in the HR diagram and eventually settled towards the Hayashi limit with a mass exceeding 600 \msolarc. This led the authors to deduce that the critical accretion rate to be approximately $1\times10^{-2}\Mpy$.

Our investigations here go beyond all previous work. By using variable accretion rates drawn from self-consistent cosmological simulations we are able 
to vary the accretion rates starting from the advent of the luminosity wave until the end of the pre-MS to obtain a more precise value of the critical accretion rate. 
Additionally, we perform numerical tests throughout the pre-MS evolution to precisely quantify this value by varying the accretion rate by hand.
We determine a value of $\rm{\dm = 2.5\times10^{-2}}\Mpy$ for the pre-MS which decreases to $\rm{\dm = 7\times10^{-3}}\Mpy$ for the Hydrogen burning phase of the stellar evolution. Our $\rm{\dot{M}_{crit,preMS}}$ of $2.5\times10^{-2}\Mpy$ is similar to the works of \citet{Omukai2001,Omukai2003,Sakurai2015,Lionel2018}. Furthermore, using model 4477 we find the existence of an additional accretion rate,  $\rm{\dot{M}_{crit}}$, during the core hydrogen burning phase which has not been explored previously in the literature. For the Hydrogen burning phase we find a value of $\rm{\dot{M}_{crit}} = 7\times10^{-3}\Mpy$.

\subsection{Numerical tests to determine \dm$_\mathrm{crit, preMS}$ and \dm$_\mathrm{crit, MS}$}

To precisely determine the critical accretion rate ($\dot{M}_{crit, preMS}$) found in this study, we performed numerical tests on model 932 throughout its pre-MS evolution. This test involved choosing constant accretion rates from $1.0 \times 10^{-3} - 3.1 \times 10^{-2} \Mpy$, and recomputing the evolution of the model from event `I' shown in the left panel of Figure \ref{Fig:HRD1}. The choice of lower limit for the accretion rate was motivated by the results of \citep{Woods2017, Lionel2018} and the upper limit was taken from our hydrodynamic simulations which is above the $\dot{M}_{crit, preMS}$ since this model migrates to red. At accretion rates of $1.0-5.0 \times 10^{-3} \Mpy$, the model contracts towards the blue, indicating the accretion timescale is longer than the surface Kelvin Helmholtz timescale.
Moving to a slightly higher accretion rate of $1.0 \times 10^{-2} \Mpy$, the model follows an oscillating behaviour and migrates between blue and red part of HR diagram until it accretes a total of 300 M$_\odot$ over 25 kyr and finally settles in the red. Increasing the accretion rate to $2.0 \times 10^{-2} \Mpy$ showed a similar oscillating behaviour but the model settled to the red in a shorter time of 12 kyrs, with a final mass of 152 M$_\odot$. Finally, the accretion was varied from $2.0 \times 10^{-2} - 3.1 \times 10^{-2} \Mpy$ and we found that at $2.5 \times 10^{-2} \Mpy$, the model directly migrates to red with a mass of 19 M$_\odot$ over 10 years, indicating this value as the $\dot{M}_{crit, preMS}$.

A similar numerical test was performed for the same model at event `IV' (later in the pre-MS) shown in the left panel of Figure \ref{Fig:HRD1} where the model migrates temporarily to blue due to a drop in accretion rate. We found that as the accretion rate dropped below the previous critical value of $2.5 \times 10^{-2} \Mpy$, the model would migrate to blue over the surface Kelvin Helmholtz timescale. Based on these tests, we conclude that critical accretion rate during the pre-MS phase is $2.5 \times 10^{-2} \Mpy$. 

We also performed such numerical tests on model 4477 during the core hydrogen burning phase to obtain a better estimate of $\dot{M}_{crit, MS}$. The choice of constant accretion rates this time ranged from $1.0 \times 10^{-6} - 2.5 \times 10^{-2} \Mpy$. We found that accretion rates below $1.0 \times 10^{-3} \Mpy$ had no effect on the position in the HR diagram as the model continued to burn hydrogen in the blue. As the accretion rate approached $4.0 \times 10^{-3} \Mpy$, the model showed an oscillating behaviour in the effective temperature from T$_{\rm eff} = 4.50 - 4.75$. Once this rate was set to $7.0 \times 10^{-3} \Mpy$, the model migrates to red over a Kelvin Helmholtz timescale and stays on the Hayshi limit until this critical rate ($\dot{M}_{crit, MS} = 7.0 \times 10^{-3} \Mpy$) is maintained. We therefore conclude that the critical accretion rate for the core hydrogen burning phase at a mass of 3984 M$_\odot$ and a central hydrogen abundance of 0.55 is $\dot{M}_{crit, MS} = 7.0 \times 10^{-3} \Mpy$. Based on physical intuition related to the thermal timescale, we suspect this accretion rate to be dependent on the mass as well as the central mass fraction of hydrogen. However, to provide quantitative answers to this question would require future study and is outside the scope of this work.

\subsection{Luminosity wave and \dm$_\mathrm{crit}$}
The existence of a luminosity wave and its relation to the expansion of protostars was first explored by \citet{Larson1972}. Using a 2 \msolar star, the author highlighted the importance of radiative transfer of entropy in a star once the temperature reaches $9.0\times10^{6}$K. This time in our study corresponds to the early pre-MS phase. The star begins to transport this radiative entropy from the central regions to the outer boundary of the core over the thermal relaxation timescale. As this wave propagates further, the star undergoes a brief expansion of radius due to the wave reaching the surface. The choice of the initial structure of a protostellar seed, whether it is assumed to be convective or radiative affects the migration of the luminosity wave from the center to the surface and furthermore has an impact on the radius as shown by \citet{Stahler1986}. \cite{Stahler1986} explored the extremely short duration of this migration (230 yr) and conclude that such a phenomenon would be difficult to observe. Our results agree with the findings of \citet{Stahler1986} and find the duration of this migration to be 190 yr. The migration of luminosity wave was subsequently studied in detail by \citet{Hosokawa_2010} in relation to accreting stars. Their choice of accretion rate was $3\times10^{-3}\Mpy$ and they concluded that once the luminosity wave is expelled from the surface, the star contracts towards the ZAMS. Further investigations \citep{Hosokawa_2013,Woods2017,Lionel2018} expanded on the range of accretion rates, finding that a critical accretion rate exists at the time when the luminosity wave breaks at the surface, which forces the star to transition to either the blue or the red. Our study confirms that the luminosity wave occurs at the very early pre-MS phase of the evolution. Secondly, all models, whether accreting or non accreting, undergo an increase in radius when the luminosity wave breaks at the surface. Thirdly, the red or blue evolution is only weakly dependent on the luminosity wave and is instead primarily affected by the accretion rate itself. Crucially, and as discussed above, there exists a critical accretion rate during the pre main sequence, \dm$_\mathrm{crit, preMS} = 2.5\times10^{-2}\Mpy$ above which the models will migrate towards red regardless of other physical processes in operation.

\subsection{Variable Accretion Rates and Model Comparisons}
A departure from constant accretion rates is expected to occur once the accretion disc around a Pop III star becomes gravitationally unstable and fragments, as shown by \citet{Stacy2010}. The migration of such fragments onto the disc and subsequently on to the star could result in a burst of accretion and give rise to a variable accretion history, see \citet{Greif2012}. The evolution of PopIII stars until the end of core silicon burning was modeled by \citet{Ohkubo2009}, who employed variable accretion rates from \citet{Yoshida2006,Yoshida2007}. These models evolved into a blue supergiant phase and concluded their helium burning stage at log(T$_{\rm eff}) > 4.6$. However, our results differ as our models transitioned to a red phase with log(T$_{\rm eff}) \approx 3.76$ at the start of helium burning and remained there until the end of their evolution. The difference is possibly due to the treatment of energy transport in the outer layers.

The evolution of many of our models with variable accretion rates is similar to models by \citet{Woods2017} and \citet{Lionel2018} with accretion rates above $10^{-2}\Mpy$, in which the models behave as red supergiant protostars. Despite varying accretion rates, \citet{Sakurai2016} found that the growth of the stellar radius that occurs in the early pre-MS phase (time $<$ 1000 years) similar to models with constant accretion rate. This result is confirmed in our models, as depicted in the right panel of Figure~\ref{Fig:Lum}, and is consistent with the findings of \citet{Sakurai2016}.

\subsection{Impact of accretion on physical parameters}
The accretion history and the episodic bursts of accretion has a significant impact on the physical parameters of models, as shown in Table~\ref{tabModels}. Although the pre-MS lifetime of all models is nearly identical for all models and follows an inverse relation to the mass, small discrepancies arise due to the accretion rate varying during this stage as seen by model 1923. Similarly, the total lifetime of the model is also affected as evident by models 1135 and 1923. This is due to the models accreting a large proportion of their final mass towards the end of their accretion history, which is longer than other models. The time spent in the red for each model increases with mass and is due to the higher net average accretion rate during the evolution. However, if the accretion rate history is erratic and above \dm$_\mathrm{crit, preMS}$, the star may spend a significant fraction of its lifetime in the red. In the case of model 1662 the star spends 59\% of it's total lifetime in the red. Finally, due to the existence of intermediate convective zones that form early in the model's evolution and its extended time in the red phase, the surface helium enrichment in this model is remarkably high (0.76).

\subsection{The Kelvin-Helmholtz Timescale versus the Accretion Timescale}\label{Sec:surfaceTKH}

The correct timescale responsible for dictating the evolutionary tracks of accreting massive stars has been explored by \citet{Stahler1986,Omukai2003,Hosokawa2009}. They describe the Kelvin Helmholtz timescale as the thermal relaxation timescale over which a star may radiate its gravitational energy ($\tau_{KH} = GM^{2}/RL$) and the accretion timescale as the characteristic timescale for stellar growth ($\tau_{accr} = \rm{M}/\dm$). The migration between blue and red follows the interplay between the two timescales; if $\tau_{accr}<\tau_{KH}$, the models will expand and migrate to the red but if $\tau_{accr}>\tau_{KH}$, they will instead undergo contraction and move to the blue. However, during the pre-MS evolution we find that some of our models (for instance model 4477 in the right panel of Figure \ref{Fig:HRD1}) continue to expand and move into the red despite $\tau_{accr}>\tau_{KH}$. Clearly something is amiss. \\
\indent This effect was first explored by \citet{Sakurai2015} and they showed that $\tau_{KH} = GM^{2}/RL$ 
is too simplistic a timescale to consider. \cite{Sakurai2015} instead invoked the Kelvin-Helmholtz timescale at the surface layer of the star. If matter is accreted sufficiently quickly onto the surface layers then the surface of the star cannot thermally relax and hence the star moves to the red - in this case it is correct to consider the timescale for that portion of the star. \cite{Sakurai2015} found that although the star may globally expand due to the offloading of matter at the surface, the interior layers of star continue to contract during this phase. The two parts of the star therefore becomes somewhat disconnected and hence using the global Kelvin Helmholtz timescale, as defined above, is incorrect. \\
\indent The Kelvin Helmholtz timescale of the outer surface layer, which takes into account the actual distribution of mass near the stellar surface is the more appropriate metric. 
\cite{Sakurai2015} define this quantity as $t_{\mathrm{KH, surf}} = \frac{f \int s_{\text{rad}} T dm}{\int dl}$, where s$_{\mathrm{rad}}$ is the entropy of radiation, T is the temperature, m is the mass of the mass of the enclosed region, l is the luminosity of the enclosed region and f is the fraction of total entropy that is carried away over the time-scale). 
\cite{Sakurai2015} applied this formula over the outer 30\% of the star (i.e. from $\rm{0.7 \ M_* \le m \le M_*}$, we will call this region the envelope). 
We also apply this expression to our model and confirm that in cases where the models expand towards the red that the surface layer Kelvin Helmholtz timescale is indeed longer than the accretion timescale i.e. $\tau_{accr}<\tau_{KH, surf}$. 
The more simplistic treatment results in  $\tau_{accr}>\tau_{KH}$ and is misleading. \\
\indent Additionally, applying the expression of the surface Kelvin Helmholtz timescale to models at the time of their transitions from red to blue (for instance, model 1662 at $\log (L/L_{\odot})=6.30$ in the right panel of Figure \ref{Fig:HRD1}), we confirm the value of the critical accretion rate, during the pre-MS ($\rm{\dot{M}_{crit, preMS}}$), derived using numerical tests to be $2.5\times10^{-2} \Mpy$ (see section \ref{Sec:surfaceKH}). At this junction $\tau_{accr} \equiv\tau_{KH, surf}$ and the critical accretion rate can be written as  $\rm{\dot{M}_{crit, preMS}} = \rm{M_{envelope}/\tau_{KH, surf}}$.

During the pre-MS, the inner regions of star are undergoing gravitational contraction while the envelope is expanding. Since $t_{\mathrm{KH, surf}}$ provides a more precise estimate of the timescale at a given Lagrangian coordinate, it is important to use this timescale to determine a red or blue transition during the pre-MS evolution. However if accretion of matter above $\rm{\dot{M}_{crit, MS}}$ were to occur during the core hydrogen burning stage, as is in case of model 4477, the interior regions of the star contract at a much longer nuclear timescale. Therefore, using the global Kelvin-Helmholtz is sufficient to predict the transition.
\section{Conclusion}\label{Sec:Conclusion}
In this study, we have investigated the evolution of massive Pop III stars under the influence of variable accretion rates. We have utilized the \gva\ stellar evolution code to model the early stages of stellar evolution and performed a detailed analysis of the critical accretion rate, the impact of the luminosity wave, and the effect of accretion on the stars' physical parameters. The major findings of this study can be summarized as follows:

\begin{itemize}
\item We have determined a critical accretion rate ($\dot{M}_{crit, preMS}$) of $2.5\times10^{-2} \mathrm{M_\odot,yr^{-1}}$ for the transition of massive Pop III stars to the red phase under disc accretion conditions during the pre-MS phase. This value is consistent with previous studies and provides a robust benchmark for future work.
\item We have found for the first time, the existence of a critical accretion rate during hydrogen burning to be $7.0\times10^{-3} \mathrm{M_\odot,yr^{-1}}$. 
\item The timescale for transitioning model 4477 from blue to red during the core hydrogen burning phase is 17.5 kyr. This timescale depends on \dm$_\mathrm{crit, MS}$ and is comparable to the global Kelvin Helmholtz timescale ($\tau_{KH} \approx GM^{2}/RL$) of the star at the time of transition.
\item We have confirmed the importance of the surface Kelvin-Helmholtz timescale in determining the transition of a star to the red or blue supergiant during the pre-MS phase, as opposed to the global Kelvin-Helmholtz timescale. 
\item The luminosity wave has been shown to only weakly impact the early pre-MS phase of stellar evolution, with the red or blue evolution being more strongly influenced by the accretion rate.
\item Our models have demonstrated that variable accretion rates have a significant impact on the physical parameters of Pop III stars, including the total lifetime, time spent in the red phase and surface helium enrichment.

\end{itemize}

In conclusion, our work contributes to the current understanding of early evolution and the complex processes governing the lives of primordial stars. The results of this study provide a solid foundation for further investigation and refinement of the processes driving the evolution of massive Pop III stars. Given that a robust determination of the critical accretion rate is hugely important for subgrid models in cosmological simulation, we advocate a value of $\dm = 7\times10^{-3} \Mpy$ in almost all instances. Where the subgrid modelling can also capture the pre-MS timescale (which will be approximately 10 kyr in duration) then the critical accretion rate for this stage should be increased to $\dm = 2.5\times10^{-2} \Mpy$.

Future work in this area could include expanding the range of accretion rates and exploring different accretion rate profiles to better understand the various factors that contribute to the observed trends. Additionally, incorporating more sophisticated models of radiative transfer, stellar winds, and mixing processes could lead to a more accurate representation of the complex processes occurring in these early stars. Ultimately, these advancements will contribute to an increasingly comprehensive understanding of the early universe and the role of primordial stars in shaping the cosmos.

\section*{Acknowledgements}

 D.N., S.E. and G.M. have received funding from the European Research Council (ERC) under the European Union's Horizon 2020 research and innovation programme (grant agreement No 833925, project STAREX). JR acknowledges support from the Royal Society and Science Foundation Ireland under grant number URF\textbackslash R1\textbackslash 191132. JR also acknowledges support from the Irish Research Council Laureate programme under grant number IRCLA/2022/1165. E.F. acknowledges support from SNF grant number 200020\_212124. TEW acknowledges support from the NRC-Canada Plaskett Fellowship. The authors would like to thank the referee for the constructive feedback and comments during the review process.

 \bibliographystyle{aa}
\bibliography{biblio}

\appendix
\section{Impact of accretion history on core hydrogen burning}\label{Sec:history}
\begin{figure*}
   \centering
    \includegraphics[width=9.0cm]{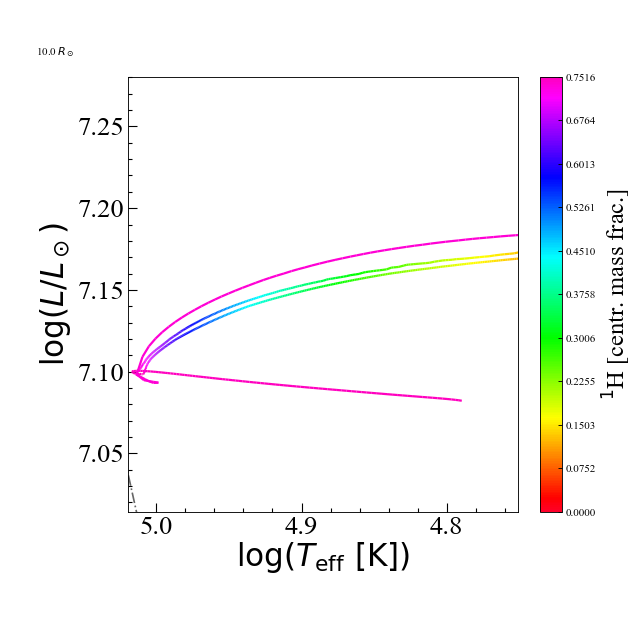}\includegraphics[width=9.0cm]{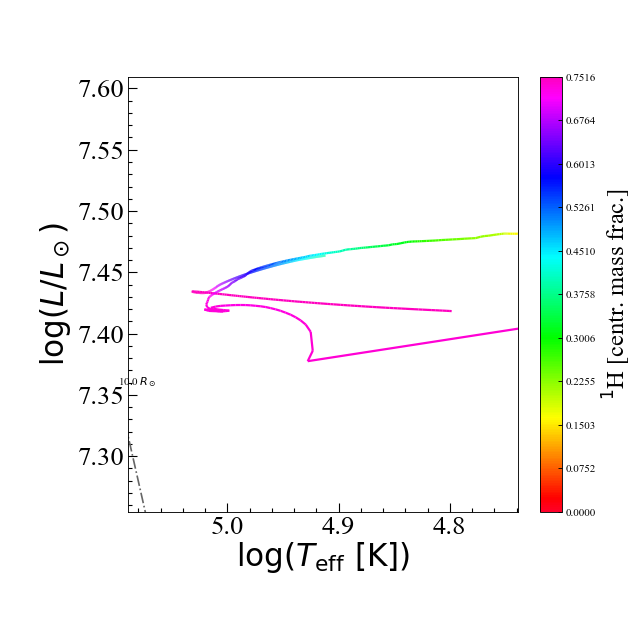}
      \caption{{\it Left panel:} A polytrope model with 491 \msolar burns hydrogen at the same place in HRD when compared to a 491 \msolar model with accretion history.{\it Right panel:} Same as left panel but for 932 \msolar.}
         \label{Fig:Tests1}
   \end{figure*}

The left panel of Figure \ref{Fig:Tests1} depicts the evolutionary track of two 491 models starting at ZAMS ($\log (L/L_{\odot})=7.10$ and $\log (T_{\rm eff})=5.08$). The second y-axis on the plot represents the central mass fraction of hydrogen. The model with a near straight migration to ZAMS is computed using a polytropic criteria with no accretion history whereas the other model represents the model 491 computed using the accretion history from hydrodynamic simulations. Despite a slight difference in the luminosity, likely due to the accreting model undergoing a pre-MS evolution where deuterium burning having a minimal impact on luminosity, the tracks are burning hydrogen in the same place at the HR diagram. This is also observed for the model 932 as shown in the right panel of Figure \ref{Fig:Tests1}. This shows that the accretion history has near negligible impact on the core hydrogen burning of a star.

\section{Accretion rates and the first 200 years of preMS evolution}
The left panel of Figure~\ref{Fig:Tests2} depicts a numerical test than explores the impact of changing accretion rates on the path length of the knee like feature observed in all accreting tracks shown in the left panel of Figure~\ref{Fig:HRD1}. In this numerical test, accretion rate at the start of computation was $1\times10^{-3}\Mpy$ and was increased by a factor of 5 every 10 years for the next 110 years. The model when compared to the 491 and 6127 models shows a near identical evolution in the early pre-MS phase of the evolution. This implies that changing accretion rate during this pre-luminosity wave phase minimally affects the path length of the knee like feature. 

\section{Impact of the choice of FITM on preMS evolution}\label{Sec:FITM}
\begin{figure*}
   \centering
    \includegraphics[width=9.0cm]{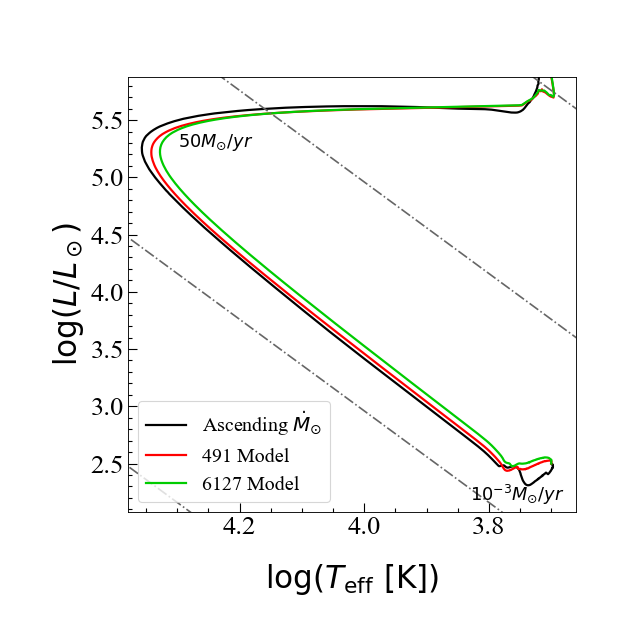}\includegraphics[width=9.0cm]{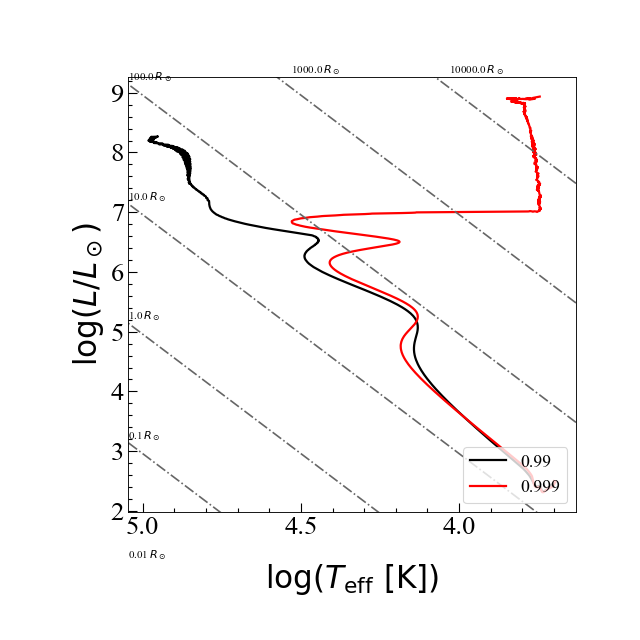}
      \caption{{\it left panel:} A zoomed-in region of the pre-MS of three models: Ascending M$_{\odot}$ (black line), 491 model (red line) and 6127 model (green line) shown in the HR diagram. For the ascending model, the accretion rate begin as $10^{-3}M_{\odot}/yr$, changes every 10 years by a factor of 5 until it reaches $50M_{\odot}/yr$ at the age of 110 years. The pre-MS evolution is very similar to 491 and 6127 models showing that at least in the early contraction phase, changing accretion rate has minimal impact on the evolutionary track. 
      {\it Right panel:}Impact of fitm on $10^{-2}$ \msolaryr models. The black line represents a FITM of 0.990 and the red line depicts a FITM of 0.999.}
         \label{Fig:Tests2}
   \end{figure*}

In \gva, FITM is defined as the mass coordinate inside a stellar model that separates the interior and the envelope and is expressed as a fraction of the actual mass. Envelope is defined as the region where the luminosity is assumed to be constant, convection proceeds non-adiabatically and partial ionisation is allowed. For instance, setting FITM at 0.999 implies the transition from interior to the envelope occurs at 0.999 of the total mass. The choice of FITM is shown to be important when exploring the accretion rates of approximately $1\times10^{-2}\Mpy$ as shown by the works of \citet{Lionel2018}. Since all 10 models computed in this study have initial accretion rates in the range of $7\times10^{-2}\Mpy$, the choice of FITM becomes important as depicted in the right panel of Figure~\ref{Fig:Tests2}. The plot depicts an accretion rate of $1\times10^{-2}\Mpy$ computed with 0.990 and 0.999. We decide to chose a FITM of 0.999 in accordance with the results of \citet{Woods2017,Lionel2018}. 

\section{The breaking of Luminosity wave}\label{Sec:Lumwave}
Figure \ref{Fig:Lum} depicts the migration of luminosity wave inside the 932 model during the early pre-MS. This wave originates right after the initial contraction and is present in all models during the pre-MS phase. The increase in central temperature reduces central opacity, transitioning the convective core to a radiative core at 12.20 \msolarc. The lowered opacity boosts luminosity production, allowing central luminosity to migrate outwards. At 18.50 \msolarc, the luminosity approaches the outer layers, as illustrated by the purple line in Figure~\ref{Fig:HRD2}. This migration, known as the luminosity wave, was described by \citet{Larson1972,Hosokawa_2010,Lionel2018}. The wave breaks at the surface when the mass reaches 19.80 \msolarc, with a luminosity of $\log (L/L_{\odot})=5.23$ and an effective temperature of log (T$_{\rm eff}) = 4.33$. All models migrate to blue region of HR diagram (log (T$_{\rm eff}) > 4.00$) 180 years into their pre-MS evolution. The breaking of luminosity wave has a weak impact on the evolution but instead, it is the accretion rate of model during and after this event that determines the migration to red or blue.

\section{Determining the $\dot{M}_{crit, preMS}$ using surface Kelvin Helmholtz timescale}\label{Sec:surfaceKH}

To better explore the impact of the surface Kelvin Helmholtz timescale on the evolution of accreting massive star models described in section \ref{Sec:surfaceTKH}, we study three instances of model 1662 and model 4477 either at or during the time of migration towards the red or the blue. 
The aim of this analytical test is to first determine the surface Kelvin Helmholtz timescale $\rm{t_{KH, surf} = \frac{\int s_{\text{rad}} T dm}{\int dl}}$ and compare it with the surface accretion timescale ($t_{\text{accr}} = \frac{\int dm}{\dot{M}}$). 
This comparison between timescales was done at $\log (L/L_{\odot})=6.3406, 7.3800, 7.4682$ corresponding to model 1662 and model 4477 as depicted in the right panel of Figure \ref{Fig:HRD1}. 
The first comparison at $\log (L/L_{\odot})=6.3406$ corresponds to the time when the impact of the critical accretion rate is first described in phase "IV" of section \ref{Sec:prems}. 
Here the values for the global Kelvin Helmholtz timescale ($\tau_{KH} = GM^{2}/RL$) for model 1662 and model 4477 are found to be 150-200 years. The global accretion timescale is instead 3600-4000 years. Upon examining the left and right panels of Figure \ref{Fig:HRD1}, we see that $\tau_{accr}>\tau_{KH}$ implies both models should migrate to blue, but instead 4477 model stays in the red. 
However, upon calculating the surface t$_{\mathrm{KH, surf}}$ and t$_{\mathrm{accr}}$ for the outer 30\% of the star, the values for the 4477 model are instead 3298 years and 1439 years respectively. 
This shows that to better estimate the blueward or redward migration of a star, it is important to calculate the timescales in the outer envelope. The results also stay consistent when applied to the outer 40 and 50\% of the star. This test was then repeated at $\log (L/L_{\odot})= 7.3800, 7.4682$ and both timescales (now much larger than before, around 30-50kyrs) follow a similar trend and provide conclusive results that support the importance of the surface Kelvin Helmholtz timescale. 

A sanity check to explore the critical accretion rate is also performed at a luminosity of $\log (L/L_{\odot})=6.3400$. Since numerical tests at this stage provide the value of $\rm{\dot{M}_{crit, preMS}}$ to be $2.5\times10^{-2} \Mpy$, we can caluclate this value again using the analytical relation above. The critical accretion rate is defined as the timescale when $\tau_{accr} \equiv\tau_{KH}$ as explained by \citet{Hosokawa_2013, Lionel2018}, we apply this value to t$_{\mathrm{KH, surf}}$ instead and obtain 1634 years and it is near identical to t$_{\mathrm{accr}}$. We can use this value and use it obtain a critical accretion rate by using the expression $\dm_{crit, preMS} = M_{current}/\mathrm{t}_{\mathrm{KH, surf}}$ and find it to be $2.52\times10^{-2} \Mpy$.

\begin{figure}
	\centering
		\includegraphics[width=8cm]{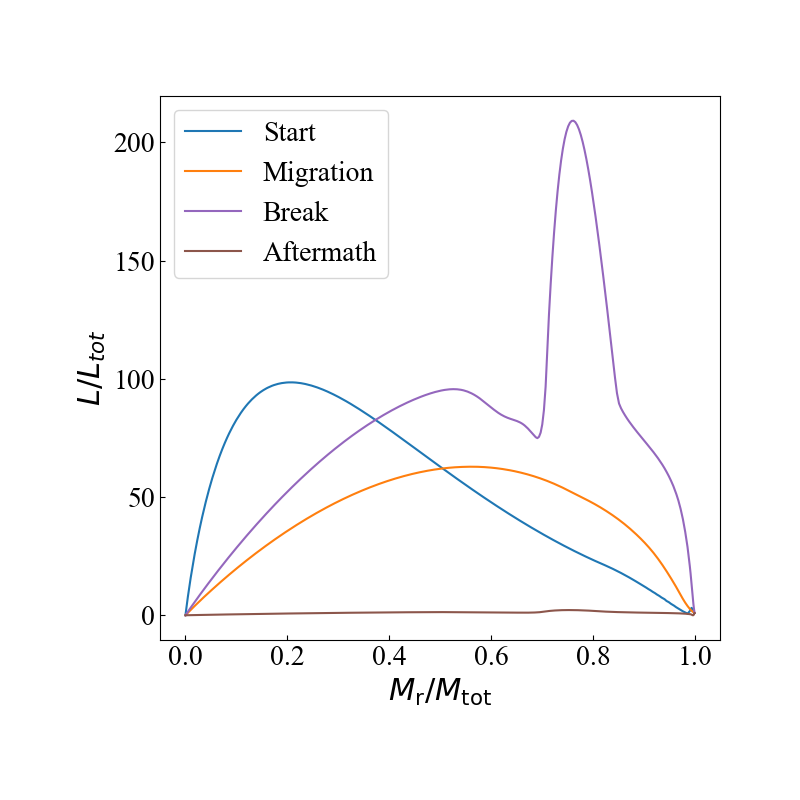}
	\caption{The propagation of luminosity wave during the pre-MS evolution of a 932 model. The wave originates deep in the stellar interior when the mass of the model is 12.20 \msolar and is shown using blue line. The outward migration of the wave is captured in orange and the model has reached 14.60 \msolarc. The wave is about to break at the surface when the mass is 18.50 \msolar and is shown in purple. If the accretion rate during the breaking of wave is higher than the critical value, the surface of the model will experience a large expansion in radius and become a red supergiant protostar. If the accretion rate is lower than the critical value, the surface of the model will expand to a much lesser extent and become a blue supergiant protostar. Once the energy is evacuated from the surface, the wave no longer exists and is shown in brown.}
		\label{Fig:Wave}
\end{figure}

\end{document}